# Integration of Radar Sensing into Communications with Asynchronous Transceivers

J. Andrew Zhang, *Senior Member, IEEE,* Kai Wu, Xiaojing Huang, *Senior Member,* Y. Jay Guo, *Fellow, IEEE,*
Daqing Zhang, *Fellow, IEEE,* and Robert W. Heath Jr. *Fellow, IEEE*

*Abstract*— Clock asynchronism is a critical issue in integrating radar sensing into communication networks. It can cause ranging ambiguity and prevent coherent processing of dis-continuous measurements in integration with asynchronous transceivers. Should it be resolved, sensing can be efficiently realized in communication networks, requiring little network infrastructure and hardware changes. This article provides a systematic overview of existing and potential new techniques for tackling this fundamental problem. We first review existing solutions, including using a fine-tuned global reference clock, and single-node-based and network-based techniques. We then examine open problems and research opportunities, offering insights into what may be better realized in each of the three solution areas.

## I. INTRODUCTION

Integrated sensing and communications (ISAC), aka, joint communications and sensing (JCAS), is a technique that enables the integration of communications and radar/radio sensing into one system, sharing a single set of transmitted signals and a majority of hardware and network infrastructure. It is considered as a major candidate in many next-generation communications systems, such as 6G mobile and Fi networks, and the next generation of radar. ISAC is expected to improve spectrum efficiency, system cost, power consumption, and performance. Integrating radar sensing into existing communication systems is known as communication-centric ISAC [1].

There are three types of geometric configurations of transmitters (Txs) and sensing receivers (Rxs) in communication-centric ISAC in terms of sensing: co-located, spatially-separated, and networked, similar to mono-static, bi-static, and multi-static/distributed radar geometries, as illustrated in Fig. 1. As discussed in [1], these configurations have respective advantages and disadvantages, and would require different levels of modifications to current communication-only networks.

The co-located mono-static configuration requires simultaneous transmission and reception operation, hence implicitly, in-band full-duplex operation, which is still immature for practical applications. Therefore, sub-optimal techniques such as deploying widely separated transmitting and receiving antennas will be needed for co-located configuration, which requires considerable system modifications [1].

The spatially-separated bi-static configuration may be a more practical near-term option for communication-centric ISAC, as it can potentially be realized without requiring any changes to the current hardware and network. However, one major issue in this configuration is the clock asynchronism, i.e., the transmitter and sensing receiver use their local oscillators with unlocked clocks. In bi-static radar, a common high-accuracy clock signal is generally provided via, e.g., fiber optical links or the GPS. However, it is typically unavailable between the base station and user terminals in current communication systems. Clock asynchronism can cause timing offset (TMO) and carrier frequency offset (CFO). Both TMO and CFO are slow time-varying due to clock instability. In communications, TMO can be absorbed into channel estimates and CFO can be estimated and compensated. In contrast, for sensing, they cause measurement ambiguity and accuracy degradation, as will be elaborated on shortly. Therefore tackling clock asynchronism is a fundamental challenge in the bi-static ISAC configuration.

In the networked configuration, multiple communication nodes can cooperatively sense the environment, such as in a cloud radio access network (CRAN). Unless a common clock is available to these nodes, the clock asynchronism issue similar to that in the bi-static configuration also exists. However, the networked environment provides more capacity for dealing with this issue, as we will elaborate on later.

In short, clock asynchronism is a fundamental problem in communication-centric ISAC. Should it be solved, bi-static and networked sensing can be efficiently realized, requiring little network infrastructure and hardware changes.

### Impact of Clock Asynchronism on Sensing

Consider a multi-input multi-output (MIMO) orthogonal frequency division multiplexing (OFDM) ISAC system. Let $H(n, t, p, q)$ be the frequency-domain channel state information (CSI) of a quasi-static channel between the $p$-th receiver antenna and the $q$-th transmitter antenna at the $n$-th subcarrier of the $t$-th OFDM block. It can be represented as

$$H(n,t,p,q) = e^{j\phi_t} \sum_{l=1}^{L} b_l \, e^{-j2\pi(\tau_l + \tau_{o,t})nf_0} e^{j2\pi(f_{D,l}+f_{o,t})tT_s} \, e^{ju_{l,p,q}}, \quad (1)$$

where $e^{j\phi_t}$ is the random phase shift term, $\tau_{o,t}$ is the TMO, $f_{o,t}$ is the CFO, $u_{l,p,q}$ is a function of angle of arrival (AoA) and angle of departure (AoD), $\tau_l$ is the propagation delay, $f_{D,l}$ is the Doppler frequency, $b_l$ is the path amplitude, $L$ is the total number of paths, and $T_s$ is the OFDM block period. In many sensing applications, from the CSI measurements, we need to explicitly or implicitly estimate $\{\tau_l, f_{D,l}\}$, as well as AoAs and AoDs, in the presence of unknown variables $\{\phi_t, \tau_{o,t}, f_{o,t}\}$ that may be slow time-varying. The CSI measurements, in the form of elements in the channel matrix $H(n,t,p,q)$, are typically obtained from channel estimation in communications, one from each packet, as shown in Fig. 1. For sensing, clock asynchronism has the following major impacts:

• TMO can directly cause timing ambiguity and hence ranging ambiguity, and CFO can cause Doppler estimation ambiguity and hence speed ambiguity. For example, for typical clock stability of 20 parts-per-million, the accumulated maximal variation of TMO over 1 millisecond can be 20 nanoseconds, which translates to a ranging error of 6 meters.

• Clock asynchronism also causes unknown and time-varying phase shifts across packets or CSI measurements, and cannot be tracked by pilots because of entangled phase shifts due to channel variation. This prevents coherently processing measurements at different timeslots/packets, and also makes Doppler estimation challenging if both the signal magnitude and phase are used.

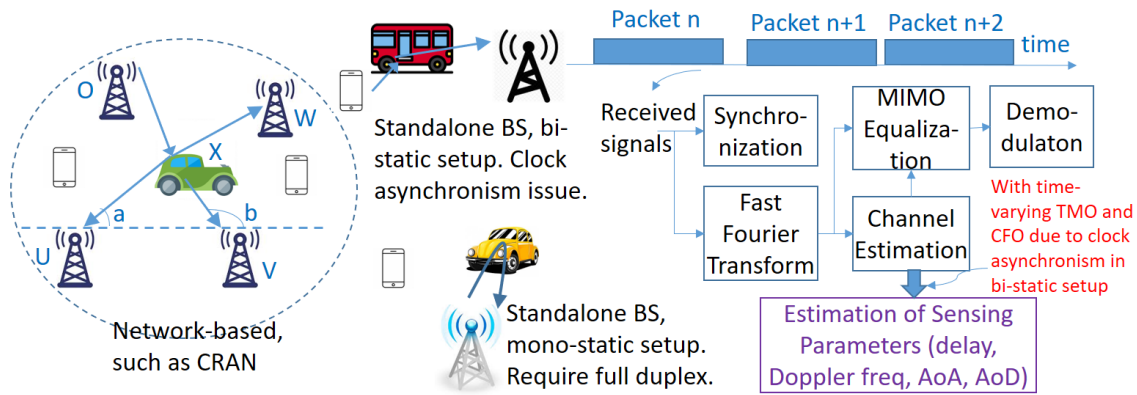

Figure 1. Three system configurations of ISAC, and a typical ISAC MIMO-OFDM receiver, where time-varying TMO and CFO due to clock asynchronism cause major sensing challenges in the bi-static setup.

| Techniques | | Merits | Issues |
|---|---|---|---|
| **Using global reference clock** | GPS disciplined | Low-cost hardware-based implementation; no additional signal processing complexity. | Require satellite visibility and be constrained to outdoor applications; Solutions that well balance synchronization speed and accuracy are yet to be developed. |
| | GPS-aided time stamping | | |
| **Single node based** | Cross-antenna cross-correlation | Exploit locked-clock across multiple receiving channels; Easy to implement without requiring changes to current network and hardware infrastructure. | Require multiple receiving channels; Constrained applications due to algorithm requirements, capability, and complexity |
| | Cross-antenna signal ratio | | |
| **Network based.** | Deterministic methods | Explore strength of networked nodes; Improved sensing capability with "multiview" and signal diversity. | Significantly increased complexity and information exchange overhead. |
| | Stochastic methods | | |

Table 1. Classification of existing solutions to the clock asynchronism problem and a brief comparison.

*Overview of Existing Solutions*

In the rest of this article, we provide a systematic review of existing sensing techniques that handle asynchronous Txs and Rxs, and look into future research directions. We classify existing techniques into three categories: using a global reference clock, single-node-based and network-based solutions, as shown in Table 1. The first relies on an external accurate reference clock. The second can be implemented in a single receiver, and the third exploits measurements from multiple cooperative nodes. Our technology review is accordingly organized into three sections next, followed by discussions on open research problems and conclusions.

## II. USING A GLOBAL REFERENCE CLOCK

The basic idea of this solution is to align individual clocks on board with a common clock source that has high reliability and stability. The evolving wireless time-sensitive network (WTSN) may be a potential solution, if its timing accuracy can be improved to the order of nanoseconds. To date, the most widely used common clock source is from the global navigation satellite system (GNSS). The standard GPS-assisted synchronization is sufficient for communications; however, further processing is required to improve the accuracy and stability of the clock signals for radar sensing applications.

Since GPS satellites broadcast time, orbit and other information via L1 and L2 signals, early time synchronization solutions receive these signals and extract the exact time information, referred to as the direct time extraction (DTE). The typical time accuracy (error between actual and estimated values) of DTE is 10 ns or less. This accuracy, however, needs to be averaged over a long time, e.g., over 1000 s, to cancel out the inherent short-term instability of satellite time. Thus, DTE can be unsuitable for applications that need fast time synchronization.

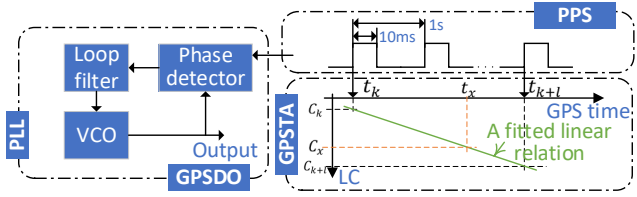

Figure 2. Illustration of PLL-based GPSDO and GPSTA.

GPS-disciplined oscillator (GPSDO) is another major solution for GPS-aided time synchronization and is widely used in distributed radar systems [2]. GPSDO is performed by aligning a radio system's local oscillator (LO) with the so-called pulse per second (PPS) signal carried by L1 and L2 signals. As illustrated in Fig. 2, PPS is a periodic pulsed signal. Thus, if distributed sensing systems are each equipped with a GPSDO, they are then synchronized to a certain extent subject to GPSDO accuracy and stability. Underlying GPSDO is generally a phased-locked loop (PLL). With the use of PLL, GPSDO suffers from the trade-off between the time constant (the time used by a PLL to lock) and lock-up performance. For distributed radars with static positions, the long waiting time for GPSDOs to synchronize may be acceptable. For ISAC sensing, however, sensing transceivers can be mobile, making the PLLs of GPSDOs frequently restart and difficult to lock.

Recently, a GPS-aided time stamping (GPSTA) method [3] was proposed to synchronize distributed wireless sensors. Unlike GPSDO, GPSTA does not change a device's LO, hence significantly reducing the time required for synchronization. In fact, GPSTA puts a timestamp on each sample, as individually performed in wireless sensors, and then resamples the digital signal to align the sampling intervals of distributed sensors, as if their clocks are synchronized. The timestamp in each sensor is estimated by mapping the local clock onto the GPS time scale, while the mapping is performed through a linear fit between the local clock count and the PPS time elapse. The basic principle of GPSTA is shown in Fig. 2. When the $k$-th PPS arrives, the absolute GPS time is extracted from the L1 signal, as denoted by $t_k$. In the meantime, the value of a local counter is recorded, as denoted by $C_k$. Similarly, when the $(k+l)$-th PPS arrives, $t_{k+l}$ and $C_{k+l}$ are recorded. Thus, if the clock count of the $x$-th sample is $C_x$, its timestamp can be estimated as $t_x = t_k + \frac{C_x - C_k}{C_{k+l} - C_k} l$. The timestamp estimation error is due to the offsets in the extracted GPS time and the reading of the local counter.

Table 2 compares the synchronization performance of the methods above. From Table2, we can conclude that GPSDO and GPSTA are more applicable to ISAC sensing than DTE due to the much faster time synchronization. Moreover, GPSDO is more suitable for static ISAC scenarios due to the PLL trade-off, while GPSTA, which is also cheaper, can be applied to both static and mobile scenarios. A major disadvantage of GPSTA is the low accuracy in terms of ISAC sensing.

| Name of Methods | Time accuracy (ns) | Frequency accuracy | Time (s) | Cost (USD) | Suitability for ISAC sensing |
|---|---|---|---|---|---|
| DTE | 3-10 (24 h) | $4 \times 10^{-14}$ (24 h) | ~1000 | -- | low |
| GPSDO | ≤5.5 | $2.6 \times 10^{-14}$ (long) | ~100 | ~1000 | Static sensing scenarios |
| GPSTA | ≤42 | -- | ~1 | ~100 | Static and Mobile |

Table 2. Performance comparison among different GPS-aided time synchronization methods.

## III. SINGLE-NODE SOLUTIONS: CROSS-ANTENNA PROCESSING

Limited research has been reported to resolve the clock asynchronism problem in a single receiver node. One set of techniques are based on constructing a reference signal from the line-of-sight (LOS) path and have been widely exploited in passive radar, such as passive coherent location [4]. The time difference of arrival (TDOA) between the reference signal and reflected echo is then measured to remove the timing offset. The technique is sensitive to the quality of the constructed reference signal. The other set of techniques commonly exploit the fact that TMO and CFO across multiple antennas in the receiver are the same, because the common oscillator clock is used in the RF circuits for all antennas. These techniques have been validated in passive WiFi sensing [5-8]. Among them, the most effective ones can be classified into two methods: cross-antenna cross-correlation (CACC), and cross-antenna signal (or CSI) ratio (CASR). Next, we mainly elaborate on the second set for their better overall performance, of which the block diagrams of some specific schemes and some experimental results are shown in Fig. 3.

### Cross-Antenna Cross-Correlation (CACC)

The CACC method computes the cross-correlation (i.e., cross-product) between signals from multiple receiving antennas. Referring to Eq. (1), we can see that CACC removes the random phase shift, TMO, and TFO; however, it outputs $L^2$ terms. The sensing parameters also become relative ones, e.g., $\tau_{l_1} - \tau_{l_2}$ and $f_{D,l_1} - f_{D,l_2}$, as well as their images $-(\tau_{l_1} - \tau_{l_2})$ and $-(f_{D,l_1} - f_{D,l_2})$.

To proceed with the estimation of sensing parameters $\{\tau_l, f_{D,l}\}$, it is widely assumed that (1) there exists a dominating LOS path with a much larger magnitude than non-line-of-sight (NLOS) paths; and (2) the transmitter and sensing receiver are static, and the relative location of the transmitter is known to the receiver. The second assumption is not necessary for some applications. Under these assumptions, the CACC outputs can be divided into four groups: the cross product of the LOS term, the cross products between NLOS terms, the cross products between LOS and NLOS terms, and the conjugates of these products. The NLOS cross products are much smaller than others and can be ignored. The LOS cross-product is invariant over the channel coherent time. Together with static terms in other groups, they can be removed by passing the CACC outputs through a bandpass filter (BPF) in the time domain. The cross-products between the dynamic NLOS and LOS paths thus dominate in the output of the filter, containing relative delays and Doppler frequencies and their images with values symmetric about zero. To this end, the outputs largely retain the linearity of the signals, if multiple paths and/or targets are present, such that conventional sensing algorithms can be easily applied. The cross-product between LOS and dynamic NLOS signals also significantly amplifies the dynamic NLOS signal, which is useful for detecting its sudden change, when, e.g., an object crosses obstacles such as walls [6]. However, one main drawback is that the image components may cause sensing ambiguity and degrade the performance of sensing algorithms.

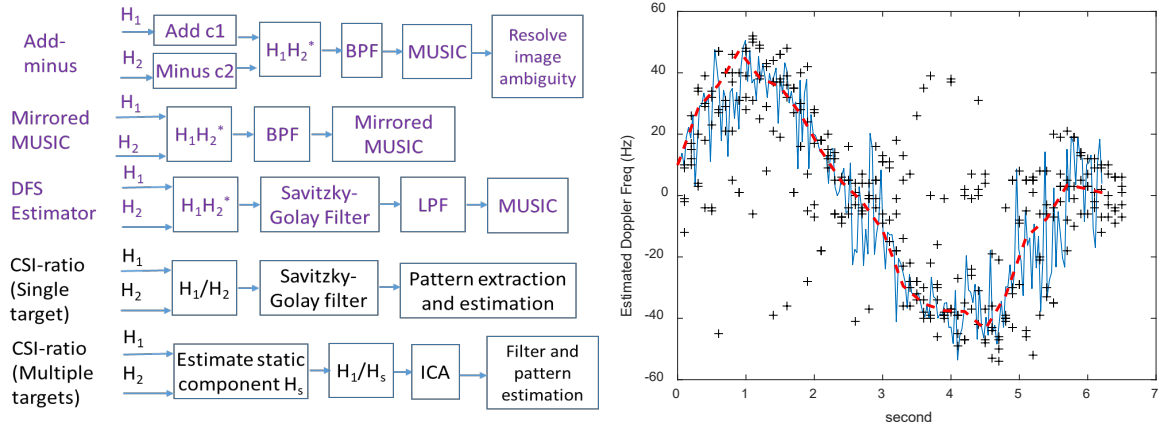

Figure 3. (Left) Block diagrams of some CACC (in purple) and CASR schemes for a system with two receiving antennas, where H1 and H2 denote the CSIs at one subcarrier of the two antennas, and the filters are applied to measurements over time for each subcarrier. (Right) Experimental results for Doppler estimation using the methods of add-minus CACC (shown in black cross marks) [5], CSI-ratio [7] (blue solid curve), and DFS estimator [10] (red dashed curve).

To suppress the image components, one widely used scheme [5], what we call *add-minus*, is as follows. Firstly, a reference antenna with typically the largest average power is selected. A constant value is then added to signals from the reference antenna, and another constant is subtracted from signals at other antennas before the cross-correlation operation. In this way, one of the two LOS-NLOS cross-products is enlarged, and the other is diminished. This scheme has been widely used for range and velocity estimation in WiFi sensing and is shown to work to some extent. However, it is found to be susceptible to the number and power distribution of static and dynamic signal propagation paths.

An alternative scheme, called mirrored MUSIC, is proposed in [9]. Exploiting the symmetry of the unknown relative parameters, we can construct two mirrored signal and basis vectors by adding the original ones in a conventional two-dimensional (2D) MUSIC algorithm with their sample-reversed versions. Thus, only $L$ mirrored vectors are needed to span the whole signal space, instead of the $2L$ ones in conventional MUSIC. This equivalently reduces the unknown parameters by half, improves the estimation performance, and resolves the image ambiguity in the outputs.

A different strategy of removing the image components is proposed in [10] for single-target real-time passive WiFi tracking. It exploits the dominating LOS cross product instead of simply removing it, unlike all the CACC methods mentioned above. The LOS cross-product is used to obtain the ratio between dynamic and static CSI. Then a metric is established as a function of the sensing parameters of the dynamic path, without involving image components. Accurate Doppler estimates can then be obtained from the metric, using, e.g., the conventional MUSIC algorithm. The signal auto-correlation is then applied to extract dynamic components by exploiting the estimated Doppler frequencies, followed by delay and AoA estimation.

*Cross-Antenna Signal Ratio (CASR)*

By exploiting the common offsets across receiving antennas, we can also compute the CSI ratios between one reference antenna and others to remove the impact of clock asynchronism [7,8], which we call the CASR technique. For a single dynamic path, CASR generates an expression where the phase associated with sensing parameters is only contained in one term in the denominator. There is a close relationship between the phase variation of the CSI ratio and that of the sensing parameters.

In [7], by considering a single dynamic path in respiration sensing, a close relationship is established between the ratio of CSI measurements across two receiving antennas and human chest movement. The CSI ratio is rewritten in the form of Mobius transformation, and the reflection path length change due to chest movement can be directly mapped to the change of CSI ratio. More specifically, it is found that the CSI ratio rotates along a circular arc clockwise or counter-clockwise, corresponding to the inhalation and exhalation in breathing pattern detection. With the application of CSI ratio, it is shown that the sensing range can be significantly extended with high accuracy [7], compared to CACC.

Using the principle of CASR, it is still possible to formulate a linear multi-source mixture problem when multiple dynamic targets need to be sensed. One example is available from [8], where respiration sensing for multiple persons is studied. This work first separates the CSI measurements for multiple persons, and then applies sensing algorithms to each person's CSI. To separate the CSIs, a blind source separation technique such as the independent component analysis (ICA) can be applied. The ICA technique requires that the sources are mixed linearly;

however, the direct CSI ratio is non-linear. Since the impact of clock asynchronism still needs to be removed before ICA, the CSI-ratio expression needs to be modified to generate a linear model. This is achieved by using the static background component as the denominator in the CSI ratio, since a common phase shift due to clock asynchronism exists in the originally received and background signals. In [8], a filter is designed through the genetic algorithm to extract static background signals from the CSI measurements. The modified ratios form linear combinations, and ICA can then be applied.

*Comparisons*

The CACC method retains signal linearity; hence conventional sensing algorithms can be easily applied to solve complicated sensing problems involving, e.g., multiple targets and multiple dynamic paths. The advantages of CASR, compared to CACC, are as follows: (1) The CSI ratio is simple to compute and has an elegant relationship with the target movement; and (2) It can cancel the distortions in CSI that are common to receiving antennas, such as the AGC variation, to improve the SINR, leading to significantly improved sensing performance such as larger sensing range. Despite these advantages, known solutions are constrained to indoor environments, considering only low-speed moving targets and estimating their relative movement, essentially associated with the Doppler frequency. Given the challenges in estimating other sensing parameters, it is unclear whether they can be applied to applications such as the characterization and tracking of high-mobility targets.

Fig. 3 presents some experimental results for Doppler frequency estimation from WiFi CSI measurements for a human walking in an office environment [5]. All estimates are raw, before smoothing and filtering operations. We can clearly see the residual images from the add-minus CACC outputs, which require complicated subsequent filtering and path matching processing to clean up [5]. The DFS estimator performs best with the smallest variation. In Table 3, the key ideas, advantages and disadvantages of these techniques are summarized.

## IV. COOPERATIVE NETWORK-BASED SOLUTIONS

In a communication network involving multiple nodes, receivers can conduct cooperative sensing, via either a deterministic or statistical approach. There has been extensive research on cooperative networked localization, which locates signal-emitting transmitters [11]. Comparatively, there has been little reported work on cooperative sensing, which aims to sense both transmitters and the environment surrounding the transmitters and receivers, and deals with both localization and diverse sensing applications. Despite these differences, many existing techniques for cooperative localization may be extended and applied to cooperative sensing. Here, we explore two classes of potential technologies to deal with clock asynchronism in cooperative sensing: deterministic geometric and statistical methods. We leave discussions for major challenges, such as target association, for cooperative networked sensing, to Section V.

*Deterministic Geometric Methods*

The deterministic geometric methods exploit known geometric relationships to remove the clock offset, via, e.g., the trilateration and triangulation techniques, which have been commonly used in networked localization [11]. Here, we depict the potential of applying these methods to networked sensing. As shown in the networked configuration in Fig. 1, multiple remote radio units (RRUs) are used to collect the echo signals from the same Tx to sense a target, e.g., the car therein. As RRUs are centrally controlled through, e.g., optical fibre, they can be well synchronized. However, there still exists clock asynchrony between Tx and RRUs.

TDOA can suppress the timing offset that is common to the RRUs. Three RRUs can result in two TDOAs with timing offset suppressed. Then, using the known locations of RRUs combined with the two TDOAs can solve the target's location unambiguously.

The AOA-based solution is relatively simpler. Only two RRUs are needed to estimate the AOAs of the same target. Then the target location can be solved using some basic triangular relations. As illustrated in Fig. 1, for the triangle XUV, if we know two angles, *a* and *b*, the location of X can be easily solved based on the known locations of U and V.

The TDOA solution needs three synchronized nodes to sense one target unambiguously. In contrast, the AOA solution can be performed over two nodes that may not be synchronized but must be equipped with antenna arrays to estimate AOAs. Since antenna arrays are commonly used in modern mobile networks, AOA-based solutions can be more promising in ISAC.

*Stochastic Methods*

Various statistical estimators have been developed for network localization, as reviewed in [13]. Some of these techniques have been explored to deal with the asynchronization problem, by exploiting the statistical averaging effect of multiple measurements.

We illustrate the methods via one example based on the expectation-maximization (EM) technique. In [14], multiple moving passive targets are localized with one Tx and multiple Rxs. RFID tags are assumed to be installed on these targets so that the reflected signals can be separated at Rxs. Thus essentially, a single passive target localization problem is considered. The receiver-specific time-varying clock offset is modelled as a memoryless Markov process where a Gaussian noise process is introduced to represent the difference between different timeslots. The Gaussian noise variables are statistically independent between Rxs, but have the same zero mean and non-zero variance parameters. This clock offset model makes it possible to average the effect of offset in the subsequently formulated maximal likelihood and maximum a posteriori estimator. The solution of the estimator is obtained by applying an iterative EM algorithm. The work demonstrates the efficiency of such a statistical estimator in handling not only the clock offset, but also NLOS links. However, the high complexity of the scheme could be a concern for real-time implementation.

| Methods | Key Idea | Advantages | Disadvantages |
|---|---|---|---|
| **Reference Signal** | Construct a reference signal from LOS path, and then compute TDOA. | Linearity of signals and flexibility of signal processing are retained. | - High-quality reference signal construction is not easy;<br>- Coherent processing of discontinuous signals is challenging due to random phase shifts. |
| **CACC** | Compute signal (CSI) cross-correlation between one reference and other antennas | - All useful signals are retained.<br>- Linearity is retained and conventional sensing techniques can be applied. | - Have application constraints;<br>- By-product cross-correlation outputs need to be removed. |
| **CASR** | Compute the signal (CSI) ratio between any two antennas | - Cancel distortions in CSI common to Rx antennas, e.g., AGC variation, to boost signal-to-interference-and-noise ratio (SINR);<br>- No image components;<br>- Easy to see phase variation. | - Hard to estimate Parameters in denominator;<br>- Linearization requires special techniques and may be hard to achieve for multiple paths/targets;<br>- CSI ratio has different characteristics from CSI, making phase estimation challenging. |

Table 3 Comparison of existing single-node cross-antenna processing techniques.

## V. FUTURE RESEARCH DIRECTIONS

Although existing solutions have demonstrated their potential in resolving the clock asynchronism problem, they have respective limitations, and there are still significant spaces for improvement. Here, we discuss open problems and research opportunities in these areas.

*Refining GPS Clocks for ISAC*

GPSTA is a promising time synchronization solution for ISAC but needs to be improved in terms of synchronization accuracy. Two potential future directions are suggested below:

*Multi-point adaptive fitting:* Rather than the two-point fitting shown in Fig. 2, one may employ multi-point adaptive fitting by continuously incorporating new GPS time and onboard counter samples. Adaptive updating algorithms can also be devised to constantly refine the time synchronization accuracy.

*Jointly using multi-satellite PPS signals:* When multiple satellites are in view, their PPS signals can be jointly used to reduce the short-term instabilities of satellite times. Note that such time instability has been unveiled as a significant time error in GPSTA [3].

*Relaxation of Single-node Sensing Techniques*

The CACC and CASR methods have demonstrated great potential, but they also have notable limitations. Tackling their limits is important for generalizing these techniques.

Firstly, the CACC method relies on the assumptions of fixed Tx-Rx locations and presence of a dominating LOS path. With varying Tx and Rx locations, the estimated sensing parameters become relative, and the absolute locations of reflectors cannot be determined. Without the dominating path, all product terms may have similar power, and the product terms with dynamic paths cannot be ignored anymore. It is critical to relax these assumptions to broaden the applications of CACC, via, e.g., exploring known static reflectors near the receiver.

Secondly, the CASR method's effectiveness has only been demonstrated for sensing relative movement. Essentially, only Doppler frequencies associated with the relative motion of targets have been estimated. The variation of propagation delay and AOAs are yet to be considered, to significantly extend the applications of CASR to scenarios involving high mobility and large range variations. Their estimation via CASR is much more challenging than estimating Doppler frequency. A potential solution would be to combine CASR with CACC.

Thirdly, extending the single-node solutions to more complicated scenarios involving multiple dynamic paths and objects will be critical for practical applications. For CASR, this is particularly challenging and only very limited works have been reported. The linear signal separation technique in [8] requires signal independence between different users, which may not always be available. It would be critical to develop more general linearization techniques for CASR so that conventional sensing algorithms can be applied to deal with these complicated scenarios. Advanced techniques based on machine learning may be applied to extract feature signals for different targets.

*Cooperative Networked Sensing*

Cooperative networked sensing based on, e.g., TDOA and AOA, can be challenging in practice due to a critical issue of target association. The issue is specific to ISAC, because, in networked localization, transmitters can be differentiated in specific domains, e.g., waveform and frequency.

In ISAC sensing, the presence of multiple targets makes the timing offset entangled with the propagation delays, and the order of path arrivals may not be the same for different receivers. Thus target association needs to be implemented before almost all timing-based trilateration operations. For triangulation, target association may be implemented simultaneously, as it is almost independent of clock offset. For timing-based trilateration, it is generally challenging, and the complexity may increase exponentially with the number of targets increasing. One potential solution is to exploit new communication protocols to assist the association [12]. The stochastic methods in networked sensing may also be adopted to realize joint association and sensing [11,13].

We envision several potential directions to combat clock asynchronism in a networked environment: 1) Instead of confining in TDOA- or AOA-based solutions from conventional localization, we can extend the degrees of freedom to more unexplored domains, such as Doppler and polarization. The combination of different domains can bring more benefits; 2) We may identify and exploit useful information from the sensing environment to perform post-calibration for all targets. The rationale is that some targets may be active users that communicate with the sensing transceivers. The clock correction can be more easily performed on these targets by cooperation. Then, the calibration information from these targets may be used for removing the clock asynchronism for other targets; and (3) we may resort to joint estimation methods, e.g., jointly estimating transmitter location and time-varying timing offset caused by clock offset and skew in [12] and AoA and frequency in [15]. These techniques typically formulate a statistically optimal objective function using, e.g., the maximal likelihood principle, and obtain ambiguity-free estimates at the cost of high complexity.

## VI. CONCLUSIONS

In this article, we show that clock asynchronism is a central problem in integrating radar sensing into communication networks, and that it can be resolved by three classes of techniques: using a global reference clock, single-node-based and network-based solutions. GPS can potentially offer a reliable global reference clock for outdoor devices, but its accuracy and stability need further improvement to meet the sensing accuracy requirement. Single-node-based techniques face the challenges of relaxing application constraints and extending application scenarios. Networked techniques need to tackle the challenging target association problems while applying trilateration and triangulation techniques. The three categories of techniques may also be combined. The prospective solutions are expected to boost the realization of integrated sensing in communications significantly.

## ACKNOWLEDGEMENTS

This research was supported partially by the Australian Government through the Australian Research Council's Discovery Projects funding scheme (project DP210101411).

## REFERENCES

1. J. A. Zhang, M. L. Rahman, K. Wu, X. Huang, Y. J. Guo, S. Chen, J. Yuan, "Enabling Joint Communication and Radar Sensing in Mobile Networks - A Survey", IEEE Communications Surveys & Tutorials, doi: 10.1109/COMST.2021.3122519.

2. W. Wang, "GPS-Based Time & Phase Synchronization Processing for Distributed SAR," in *IEEE Trans. on Aerospace and Electronic Systems*, vol. 45, no. 3, pp. 1040-1051, 2009

3. Koo KY, Hester D and Kim S (2019) Time Synchronization for Wireless Sensors Using Low-Cost GPS Module and Arduino. Front. Built Environ. 4:82.

4. H. Kuschel, D. Cristallini and K. E. Olsen, "Tutorial: Passive radar tutorial," in IEEE Aerospace and Electronic Systems Magazine, vol. 34, no. 2, pp. 2-19, 2019.

5. K. Qian, C. Wu, Y. Zhang, G. Zhang, Z. Yang and Y. Liu, "Widar2.0: Passive Human Tracking with a Single Wi-Fi Link", Proc. International Conf. Mobile Systems, Applications, and Services, pp. 350-361, 2018

6. S. Li, Z. Liu, Y. Zhang, Q. Lv, X. Niu, L. Wang, and D. Zhang. WiBorder: Precise Wi-Fi based Boundary Sensing via Through-wall Discrimination. Proc. ACM Interact. Mob. Wearable Ubiquitous Technol. vol. 4, no. 3, 2020.

7. Y. Zeng, D. Wu, J. Xiong, E. Yi, R. Gao, and D. Zhang, "Farsense: Pushing the range limit of wifi-based respiration sensing with CSI ratio of two antennas," Proc. ACM Interact. Mob. Wearable Ubiquitous Technol., vol. 3, no. 3, 2019.

8. Y. Zeng, D. Wu, J. Xiong, J. Liu, Z. Liu, and D. Zhang, "Multisense: Enabling multi-person respiration sensing with commodity WiFi," Proc. ACM Interact. Mob. Wearable Ubiquitous Technol., vol. 4, no. 3, 2020.

9. Z. Ni, J. A. Zhang, X. Huang, K. Yang and J. Yuan, "Uplink Sensing in Perceptive Mobile Networks with Asynchronous Transceivers," in IEEE Trans. on Signal Processing, vol. 69, pp. 1287-1300, 2021

10. Z. Wang, J. A. Zhang, M. Xu, and Y. J Guo, "Single-Target Real-Time Passive WiFi Tracking", IEEE Trans. On Mobile Computing, doi: 10.1109/TMC.2022.3141115.

11. C. Laoudias, A. Moreira, S. Kim, S. Lee, L. Wirola and C. Fischione, "A Survey of Enabling Technologies for Network Localization, Tracking, and Navigation," in IEEE Communications Surveys & Tutorials, vol. 20, no. 4, pp. 3607-3644, 2018

12. J. Zheng and Y. -C. Wu, "Joint Time Synchronization and Localization of an Unknown Node in Wireless Sensor Networks," in IEEE Trans. on Signal Processing, vol. 58, no. 3, pp. 1309-1320, 2010

13. H. Wymeersch, J. Lien and M. Z. Win, "Cooperative Localization in Wireless Networks," in Proceedings of the IEEE, vol. 97, no. 2, pp. 427-450, 2009

14. W. Yuan, N. Wu, B. Etzlinger, Y. Li, C. Yan and L. Hanzo, "Expectation–Maximization-Based Passive Localization Relying on Asynchronous Receivers: Centralized Versus Distributed Implementations," in IEEE Trans. on Communications, vol. 67, no. 1, pp. 668-681, 2019

15. K. Hsu, and J. Kiang. "Joint Estimation of DOA and Frequency of Multiple Sources with Orthogonal Coprime Arrays." Sensors 19.2 (2019): 335.

## BIOGRAPHIES

**J. Andrew Zhang** (Andrew.Zhang@uts.edu.au, Senior Member, IEEE) is an associate professor at the University of Technology Sydney (UTS). He received his PhD from the Australian National University in 2004.

**Kai Wu** (Kai.Wu@uts.edu.au) is a research fellow with UTS. He received his PhD from UTS in 2020.

**Xiaojing Huang** (Xiaojing.Huang@uts.edu.au. Senior Member, IEEE) is a professor with UTS. He received his PhD from Shanghai Jiaotong University in 1989.

**Y. Jay Guo** (Jay.Guo@uts.edu.au, Fellow, IEEE) is a distinguished professor with UTS. He received his PhD from Xi'an Jiaotong University in 1987.

**Daqing Zhang** (dqzhang@sei.pku.edu.cn, Fellow, IEEE) is a professor with Peking University, China, and Telecom SudParis, IP Paris, France. He received his Ph.D. from University of Rome "La Sapienza", Italy in 1996.

**Robert Heath** (rwheathjr@ncsu.edu, Fellow, IEEE) is a professor with North Carolina State University, USA. He received his Ph.D. from Stanford University, Stanford, CA, in 2002.